\documentclass[prb,twocolumn,amsmath,amssymb,superscriptaddress]{revtex4}
\usepackage[T1]{fontenc}
\usepackage[latin1]{inputenc}
\usepackage{graphics}
\usepackage{amssymb}
\usepackage{amsmath}
\usepackage{overpic}

\newcommand{\be}{\begin{equation}}
\newcommand{\ee}{\end{equation}}
\newcommand{\bea}{\begin{eqnarray}}
\newcommand{\eea}{\end{eqnarray}}

\def\a{\alpha}
\def\b{\beta}
\def\e{\varepsilon}
\def\d{\delta}

\def\l{\lambda}

\def\o{\omega}

\def\s{\sigma}

\def\G{\Gamma}

\def\O{\Omega}

\def\ra{\rightarrow}

\def\pd{\partial}

\def\bk{{\bf k}}

\def\nn{\nonumber}
\def\lb{\label}
\def\pref#1{(\ref{#1})}

\newcount\bozza \bozza=0
\ifnum\bozza=1
\newdimen\shift \shift=-2truecm
\def\lb#1{%
{\label{#1}\rlap{\kern\shift{$\scriptstyle#1$}}}}
\else\def\lb#1{\label{#1}} \fi

\begin{document}

\title{Effects of the Fermi-surface shrinking on the
optical sum rule in pnictides}

\author{L. Benfatto}

\affiliation
{Institute for Complex Systems (ISC), CNR, U.O.S. Sapienza and \\
Department of Physics, Sapienza University of Rome, P.le A. Moro 2,
00185 Rome, Italy}

\author{E. Cappelluti}
\affiliation
{Institute for Complex Systems (ISC), CNR, U.O.S. Sapienza and \\
Department of Physics, Sapienza University of Rome, P.le A. Moro 2,
00185 Rome, Italy}

\date{\today}

\begin{abstract}
  In this paper we investigate the effects of the band shifts induced
  by the interband spin-fluctuation coupling on the optical sum rule
  in pnictides.  We show that, despite the shrinking
  of the Fermi surfaces 
  with respect to first-principle calculations, the charge-carrier concentration
  in each band is almost unchanged, with a substantial conservation of
  the total optical sum rule. However, a significant transfer of
  spectral weight occurs from low-energy coherent processes to
  incoherent ones, with practical consequences on the experimental
  estimate of the sum rule, that is carried out integrating the data
  up to a finite cut-off. This has profound consequences both on the
  absolute value of the sum rule and on its temperature dependence,
  that must be taken into account while discussing optical experiments
  in these systems.
\end{abstract}


\maketitle

\section{Introduction}

Since the discovery of superconductivity in iron-based superconductors
a renewed interest emerged on the properties of interacting multiband
systems. Indeed, all the families of pnictides are semimetals, with
several small hole and electron pockets at the Fermi level,
originating from almost empty hole and electron bands. Such a topology
has been predicted by Density Funcional theory
(DFT) calculations and confirmed by several
experiments sensitive to the Fermi-surface structure, as de Haas-van
Alphen\cite{coldea,analytis,carrington_prl10} and angle-resolved
photoemission spectroscopy (ARPES)\cite{lu,yang,wray,yi}.
Despite the qualitative agreement with DFT predictions as far as the
number and the character of the bands is concerned, the experimental data
suggest the existence in the real materials of a band narrowing
operating over an energy scale of hundredths meV and of a shrinking of
the Fermi-surface areas with respect to DFT.

The first feature has been associated to the presence of electronic
correlations\cite{georges,vollhardt} leading to an enhancement of the
band mass $m$ of the carriers with respect to the DFT value $m_{\rm
  DFT}$, in agreement with several DFT+DMFT calculations giving
$m\simeq 2 m_{\rm DFT}$.  This effect can be possibly detected also by
an experimental estimate of the optical sum rule, which relates the
integrated optical conductivity $\s(\o)$ to the carrier density $n$
and band mass, $W =\int_0^\infty d\omega \sigma(\omega)\propto
n/m$.\cite{review_benfatto, review_basov} Indeed, recent experiments in LaFePO show that the experimental
weight $W_{\rm exp}$ is significantly reduced with respect to the DFT
value, $W_{\rm exp}/W_{\rm DFT}\approx 0.5$.\cite{basov} Assuming that
DFT correctly estimates the number of carriers $n$ in each band, this
result can be interpreted thus as an effect of the band-mass
renormalization mentioned above, $W_{\rm exp}/W_{\rm DFT}\sim
m_{\rm DFT}/m$.

In addition to the electronic correlations, operating on large energy
scales, a retarded low-energy interaction, mediated by the
spin-fluctuations, is also present in pnictides.\cite{kuroki,chubukov}  Within this context,
the shrinking of the Fermi-surface areas has been 
explained as an effect of the band shifts
originating from the interband coupling of electron and hole bands
mediated by low-energy spin-fluctuations.\cite{cappelluti_dhva}
It is worth noting that
these Fermi-surface shrinkings do not violate the total charge
conservation,  which, in multiband systems, is determined  by the balance
of the charge carriers in {\em all} the bands.
In contrast to the many-body picture, an alternative explanation for
the Fermi surface shrinkings has been proposed to
be an intrinsic inaccuracy of the DFT
calculations in determining the top/bottom energy of the bands with
respect to the Fermi level.\cite{mazin_cm} In this case, a simple
rigid shift of the DFT bands could account for the observed shrinking
of the Fermi-surface areas\cite{coldea} without involving many-body
interaction effects.  A natural consequence of this shift
would be a reduction of the carrier concentration in real materials
with respect to the DFT prediction, with direct implications
for the optical sum rule. Indeed, the experimental observation of a
low value of  $W_{\rm exp}$ could not be attributed only to
mass-renormalization effects, but should be associated (at least in
part) to a reduction of $n$ due to the
rigid-band shift.\cite{mazin_cm,basov_cm}

\begin{figure}[t]
\includegraphics[scale=0.3, angle=0,clip=]{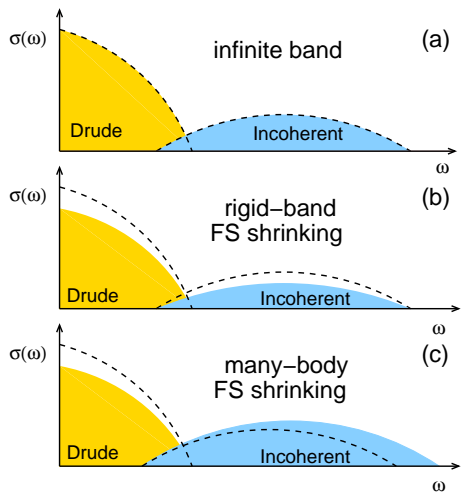}
\caption{(color online) Sketch of the
optical conductivity in three representative cases:
(a) infinite-band system interacting with a bosonic mode.
The optical conductivity is characterized by a coherent Drude peak
and an incoherent band.  No Fermi-surface shrinking is present;
(b) same as in (a), but with a Fermi-surface shrinking due to a rigid band
shift.
In this case both the Drude peak and the weight of the incoherent
processes are reduced;
(c) effects of a Fermi-surface shrinking
due to a many-body interaction in a multiband system. In this case the
total spectral weight is (almost) unchanged, but the Fermi-surface
shrinking reflects in a reduction of the coherent
spectral weight in the Drude part, with an additional increase of the
incoherent processes.}
\label{f-skt}
\end{figure}

Motivated by this framework, in this paper we present an extensive
analysis of the effect of the Fermi-surface shrinking induced by the
interband interactions on the charge-carrier conservation and on the
optical properties of a multiband system.  In particular we show
that within this context, in contrast to a Fermi-surface shrinking
induced by rigid-band shift,\cite{mazin_cm} the charge-carrier
concentration in each band does {\em not} scale with the Fermi
area,  and it is instead only weakly affected by its shrinking,
leading to a redistribution of spectral weight between coherent and
incoherent processes and to a substantial conservation of the optical
sum rule with respect to the non-interacting case.  Notably, as we
will show, the redistribution of optical spectral weight is remarkably different
with respect to the rigid band shift suggested in
Ref. [\onlinecite{mazin_cm}], where the reduction of the Fermi area is
reflected in a corresponding reduction of {\em both} the Drude peak
and the incoherent band (Fig. \ref{f-skt}).  On the contrary, in the
case of a Fermi-surface shrinking induced by the retarded interband
interactions, the reduction of the coherent Drude part associated to
the Fermi-surface shrinking is accompanied by an additional transfer of
spectral weight to incoherent processes (Fig.\ \ref{f-skt}c). As we
shall see, this has practical consequences on the experimental
determination of the optical sum rule, which is usually carried out
integrating the data up to a finite cut-off.

The structure of the paper is the following. In Sec. II we introduce
the model and we show how in a multiband system with interband
interactions the charge-carrier concentration in each band
does not scale with the
corresponding Fermi-surface area. In Sec. III we compute the optical
conductivity in the presence of interactions and we analyze the
effects of the Fermi-surface shrinking on the optical spectra. In
Sec. IV we focus on the temperature effects on the
optical-conductivity sum rule, by considering a bosonic spectrum that
is either constant in temperature or temperature dependent. In Sec. V
we comment on the outcomes of our results for the experiments in
pnictides, and in Sec. VI we summarize our conclusions.

\section{Fermi-surface shrinking and number of carriers}

In this paper we investigate the effect of a retarded interaction on
the optical properties of a multiband system characterized by two
peculiar features: (i) a strong particle-hole asymmetry of the bands
and (ii) a predominant interband character of the interaction.  The
outcomes of these features on the one-particle properties have been
discussed in Refs. [\onlinecite{cappelluti_dhva,benfatto_prb09}], and
the main results relevant to the present analysis will be recalled
below. We will then follow here the same effective low-energy
multiband scheme of 
Refs. [\onlinecite{cappelluti_dhva,benfatto_prb09}], which contains also
the main ingredients needed to discuss the physics of iron-based
superconductors. In particular, we will focus on the parameter values
appropriate for LaFePO, where both de Haas-van Alphen measurements of
the Fermi-surface shrinking\cite{coldea} and measurements of the
optical conductivity\cite{basov} are available in the literature.

\begin{figure}[t]
\includegraphics[scale=0.9, angle=0,clip=]{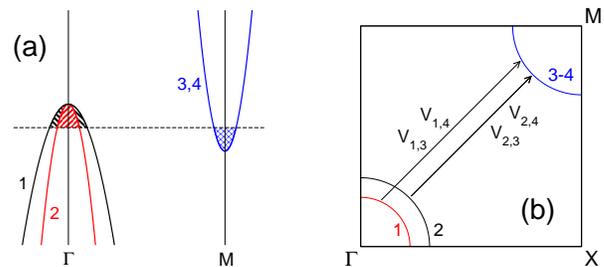}
\caption{(color online) Sketch of the band structure and of the
Fermi surfaces with the main
interband interactions.}
\label{f-bands}
\end{figure}

We consider two hole-like bands ($1,2$) located close to the $\Gamma$
point and two electron-like degenerate bands ($3,4$) close to the M
point (see Fig. \ref{f-bands}).
All the bands close to the Fermi level
can be schematized as parabolic,
respectively
\be
\lb{disph}
\epsilon_{{\bf k},\a}=
E_{{\rm max},\a}-\frac{\hbar^2|{\bf k}|^2}{2m_\a}, \quad {\a=1,2},
\ee
for the hole bands and
\be
\lb{dispe}
\epsilon_{{\bf k},\a}=
E_{{\rm min},\a}+\frac{\hbar^2|{\bf k}|^2}{2m_\a}, \quad {\a=3,4},
\ee
for the electron ones. 
We assume also that carriers
interact via a bosonic mode with a typical energy scale $\omega_0$. In the
case of pnictides, where the largest source of interactions comes from the
exchange of spin fluctuations between quasi-nested hole and electron
pockets, the interactions $V_{\a\b}$ between two bands $\a$ and $\b$ will
have a predominant interband character. As it has been discussed in Ref.\
[\onlinecite{benfatto_prb09}], the coupling to a low-energy mode cannot
account for the overall reduction of the bandwidth with respect to DFT
observed in ARPES experiments. This effect must be attributed to Coulomb
repulsion occurring at high-energy scales,\cite{georges,vollhardt}
and we model it by assuming our input masses
as a factor twice larger than the DFT estimate,
$m_\a\approx 2 m_{\a,{\rm DFT}}$.\cite{cappelluti_dhva,benfatto_prb09}
To be more explicit,
setting for convention the Fermi level $\mu=0$, we first
estimate from DFT calculations
the Fermi vectors $k_{\rm F,\alpha}$
and the nearest band edge for each band, namely
$E_{\rm max,1}$, $E_{\rm max,2}$ for the hole-like bands, $E_{\rm min,3}$,
$E_{\rm min,4}$ for the electron ones. From these we estimate the
non-interacting mass for each band $m_\alpha=\hbar^2
k_{\rm F,\alpha}^2/2E_{\mathrm{max(min)},\alpha}$ and the corresponding bare
density of states $N_\alpha^0=m_\alpha a^2/2\pi \hbar^2$
($a$ is here the in-plane
lattice constant). Accordingly, the effective band edge far from the Fermi
level \cite{note-edge} follows from the relation $N_\alpha^0=1/(E_{\rm
max,\alpha}-E_{\rm min,\alpha})$. Finally, to account for the correlation
effects we simply divide by 2 the values of the band edges extracted by
DFT, which corresponds to the mentioned
increase of a factor 2 for the effective mass $m_\a$ and for
the density of states $N^0_\a$.
The estimated band parameters in the presence of correlation
are summarized in Table \ref{t-table}. 

\begin{table}[t]
\begin{center}
\begin{tabular}{ccccc}
\hline \hline
band & 
$m_\alpha/m_e$ & $E_{\rm max,\alpha}$ & $E_{\rm min,\alpha}$
& $N^0_\alpha$   \\
index&  & (eV) & (eV) &  (eV$^{-1}$) 
\\
\hline
1   & 1.16 & 0.102 & -2.516 & 0.382 \\ 
2   & 2.28 & 0.102 & -1.231 & 0.750  \\
3,4 & 1.58 & 1.776 & -0.147 & 0.520  \\
\hline \hline
\end{tabular}
\end{center}
\caption{Microscopic band parameters used in this work. Following the
  approach of Ref.\ \onlinecite{cappelluti_dhva}, we use DFT band parameters
  renormalized by a factor 2, to account for the correlation effects.}
\label{t-table}
\end{table}

In the presence of a retarded boson-mediated interaction,
the Green's function $G_\alpha(z)$ for a generic band $\a$
can be written as:
\begin{eqnarray}
\lb{green}
G_\alpha(\bk,\omega)
&=&
\frac{1}{\omega-\epsilon_{{\bf k},\a}-\chi_\alpha(\omega)+\mu+
i\Gamma_\alpha(\omega)},
\end{eqnarray}
where we introduced the short-hand notations 
$$\chi_\alpha(\omega)={\rm Re} \Sigma_\a(\omega), \quad
\Gamma_\alpha(\omega)=-{\rm Im} \Sigma_\a(\omega).$$
The self-energy $\Sigma_\a(\omega)$ in the Matsubara space
can be computed as
\begin{eqnarray}
\lb{self}
\Sigma_\alpha(i\omega_n)
&=&
-T\sum_{m,\beta}
V_{\alpha,\beta}
D(\omega_n-\omega_m)
G_\beta(i\omega_m),
\end{eqnarray}
where 
$G_\beta(i\omega_m)=\int (d^2{\bf k}/4\pi^2)G_\beta({\bf
  k},i\omega_m)$ 
is the local one-particle Green's function,
$D(\omega_l)= \int d\Omega 2\Omega B(\Omega)/(\Omega^2+\omega_l^2)$ 
is the propagator of the bosonic mode and $B(\Omega)$ is the density
of states of the bosonic excitations. Here we assumed for simplicity
that the interaction does not depend on the momentum, so that also the
self-energy is momentum-independent.  Finally, the self-energy
$\Sigma(\omega)$ and the Green's function $G_\alpha(\bk,\omega)$ on
the real-frequency axis can be obtained employing the standard
Marsiglio-Schossmann-Carbotte analytical continuation\cite{msc} and
the number of charge carriers per band can be obtained as
\bea
\lb{defdensh} n_\a &=&
N_s\int d\o [1-f(\o)] N_\a(\o)  \quad \a=1,2\nn\\
\eea
for the hole bands and
\bea
\lb{defdense}
n_\a
&=&
N_s\int d\o f(\o) N_\a(\o)  \quad \a=3,4\nn\\
\eea 
for the electron bands, 
where $f(x)=1/[\exp(x/T)+1]$ is the Fermi function, $N_\a(\o)$ is the
interacting density of states (DOS), $N(\o)=N_\a^0\int d\epsilon A(\epsilon,\omega)$, and
where $A(\epsilon_{\bf k},\omega)=-(1/\pi)\mbox{Im}G({\bf
  k},\omega+i0^+)$ is the one-particle spectral function.

To account for a
spin-mediated interaction mechanism we use here the Lorentzian
spectrum\cite{millis_prb92}  
\be
\lb{mmp}
B(\O)=\frac{1}{\pi}\frac{\O\o_0}{\o_0^2+\O^2}
\ee
with a characteristic energy scale $\o_0=20$ meV.  From the above
relations we can introduce the matrix of the dimensionless coupling
$\lambda_{\a,\b}=V_{\a,\b}N_\b D(0)$, which is related to the
low-energy mass renormalization due to the retarded interaction as
$m_\a^*=(1+\lambda_\a)m_a$, where $\lambda_a=\sum_\b
\lambda_{\a,\b}$. Following Ref.\ \onlinecite{cappelluti_dhva} we take
$V_{\a\b}=V$ for the $(\a,\b)$ values shown in Fig. \ref{f-bands}, and
$V_{\a\b}=0$ otherwise. This choice is appropriate for LaFePO, while
for 122 pnictides the anisotropy of $V_{\a\b}$ must be taken into
account.\cite{benfatto_prb09} Nonetheless, all the results we will
discuss below depend only quantitatively on the exact form of the
spectrum and of the $V_{\a\b}$ matrix, and will be qualitatively the
same also for other choices of the boson mediator and of the
interactions, provided that interband terms $V_{\a\b}, \a\neq \b$
dominate over the intraband ones.

\begin{figure}[t]
\includegraphics[scale=0.3,angle=0,clip=]{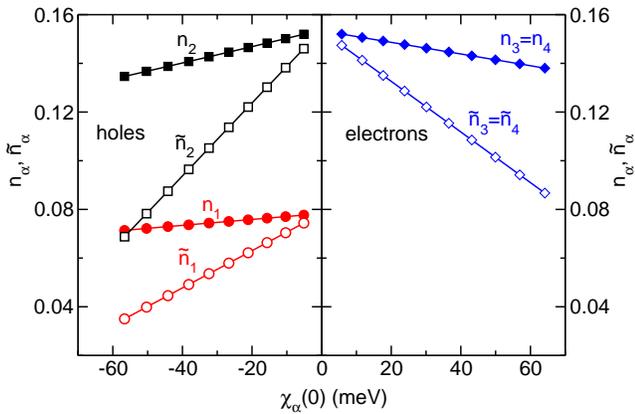}
\caption{(color online)
Variation of the hole and electron densities as a function of the
energy shift at the Fermi surface. Here $n_\alpha$ is the total number of
carriers in each band, while $\tilde n_\a$ is the number of coherent
carriers related to the Fermi area according Eq. (\ref{nshift})}
\label{f-dens-vs-chi}
\end{figure}

As we discussed in Ref. \onlinecite{cappelluti_dhva}, the strong
particle-hole asymmetry of each band in pnictides forces us to
calculate the Eliashberg self energy taking into account the finite
bandwidth.  In contrast to the usual infinite-band Eliashberg
approximation, this leads to a finite value of
$\chi_\alpha(\omega=0)$, and to a corresponding change of size of the
Fermi surface which depends on the interband $vs$. intraband character
of the interaction.  In the case of dominant interband scattering, in
particular, these finite-bandwidth self-energy effects result in a
{\em shrinking} of the Fermi surface, whose new Fermi vectors can be
obtained from the poles of the Green's function \pref{green} as 
\be
\lb{kfh} 
k_{{\rm F},\a}^2=\frac{2m_\a \left[E_{{\rm
        max},\a}+\chi_\a(0)\right]}{\hbar^2}, 
\ee 
for hole bands
$(\a=1,2)$, while for electron bands $(\a=3,4)$ we have 
\be 
\lb{kfe}
k_{{\rm F},\a}^2=\frac{2m_\a \left[|E_{{\rm
        min},\a}|-\chi_\a(0)\right]}{\hbar^2}.  
\ee 
Note that, as shown in Ref.\ \onlinecite{cappelluti_dhva}, for interband
interactions $\chi_\a(0)$ is {\em negative} for the hole bands while
it is {\em positive} for the electronic one, so that $k_{{\rm F},\a}$
is reduced in both cases.  From the Fermi area we can define a {\em
  coherent} charge-carrier concentration per band: 
\be 
\lb{nshift}
\tilde{n}_\a=\frac{a^2 k_{{\rm F},\a}^2}{2\pi}, 
\ee 
where we have
taken into account the spin degeneracy $N_s=2$.  In the case of a
rigid band shift, as hypothesized in Ref. \onlinecite{mazin_cm},
$\tilde{n}_\a$ is also equivalent to the total charge-carrier
concentration per band $n_\a$.
Things are
however deeply different when the shrinking of the Fermi surface is
induced by a retarded interaction mediated by a low-energy boson.
Indeed, in this case although the reduction of the Fermi areas and
hence of $\tilde{n}_\a$ is remarkable, the total charge carrier
concentration per band $n_\a$ is almost unaffected.  Note that this
result does not violate the Luttinger theorem, since for a multiband
system there is no equivalence within each band between the charge
concentration $n_\a$ and its corresponding Fermi area encoded in
$\tilde n_\a$, as it is instead the case in a single-band system.
Indeed, the usual argument used to prove such an equivalence for the
single-band case\cite{luttinger60} shows that in the multiband case the Luttinger
theorem only applies to the sum of the carrier density in all the
bands. In
Fig.\ \ref{f-dens-vs-chi} we plot the variation of both the total
$n_\a$ and the coherent $\tilde{n}_\a$ carrier density for each band
as a function of the corresponding band shift $\chi_\a(0)$, which is
 proportional to the coupling to the bosonic
mode.\cite{cappelluti_dhva} The corresponding shrinking of the Fermi
area can be directly obtained by Eqs. (\ref{kfh})-(\ref{kfe}).  As one
can see, a band shift $\chi_\a(0) \approx 25$ meV (that corresponds to
the Fermi-surface shrinking observed experimentally in
LaFePO\cite{coldea,cappelluti_dhva}) leads to a remarkable reduction
of $\tilde{n}_\a$, up to almost 25$\%$, whereas the reduction of
$n_\a$ is much less pronounced, with largest variations of $n_\a$ of
the order of 5\%.

\begin{figure}[t]
\includegraphics[scale=0.3,angle=0,clip=]{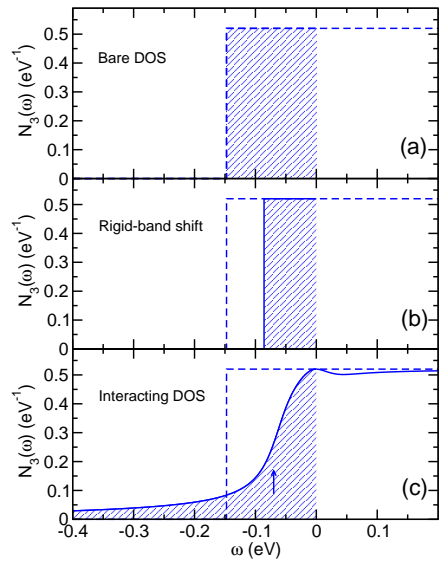}
\caption{(color online)
DOS of the electron band 3 in three representative cases: (a) 
bare DOS; (b) rigid-band shift; (c) interacting case,
computed using $V=0.92$ eV. The shaded areas correspond to the
particle number at $T=0$ in the three cases. The arrow in panel (c)
marks the new band edge due to the energy shift induced by
$\chi_3(0)$. Note however that the DOS extends also below this limit due to the
finite value $\G_3(\o)$ of the imaginary part of the self-energy.}
\label{f-dos}
\end{figure}

In order to understand why the total charge-carrier concentration per
band $n_\a$ is almost unaffected by the interaction in spite of the
strong shrinking of the Fermi areas, let us examine the behavior of
the interacting DOS $N(\o)$.  As an explicative example, we compare in
Fig. \ref{f-dos} the DOS of the electron band 3 in three cases: (a)
without interaction, (b) in the presence of a rigid-band shift
equivalent to $\chi_3(0)$, and (c) in the presence of the coupling to
spin fluctuations. Here we used a relatively large value of the
coupling ($V=0.92$ eV, leading to $\chi_3(0)\approx 60$ meV) to make more
visible in the figure the effects of the interaction. In each panel
the shaded area corresponds to the integral of the DOS up to Fermi
level (i.e.  $\o=0$ in our notation), which gives the carrier density
at $T=0$ according to Eq.\ \pref{defdense}. In the case of a
rigid-band shift the number of carriers decreases following the
Fermi-surface shrinking \pref{nshift}. However, as shown in Fig.\
\ref{f-dos}c, when the shrinking is due to a many-body interaction the
redistribution of spectral weight in the DOS is remarkably
different. Here the presence of a finite $\chi_3(\o)$ is reflected, in
analogy with the rigid-band case, in a shift of the band bottom with
respect to the non-interacting case, visible at the energy level where
the quasi-particle DOS has a rapid drop (blue arrow in Fig.\
\ref{f-dos}c).  However, this effect is balanced by a strong
redistribution of spectral weight in an extended tail of the DOS for
$\omega \lesssim E_{\rm min,3}=-0.15$ eV, which (almost) compensates
the band-edge shift and explains why $n_3$ is only slightly smaller
than the bare value, as shown Fig.\ \ref{f-dens-vs-chi}.  Such long tail in
the DOS is due to the incoherent states induced by the imaginary part
of the self-energy $\G_3(\o)$, and it is a characteristic signature of
the interband nature of the interaction. 

\begin{figure}[t]
\includegraphics[scale=0.4, angle=0,clip=]{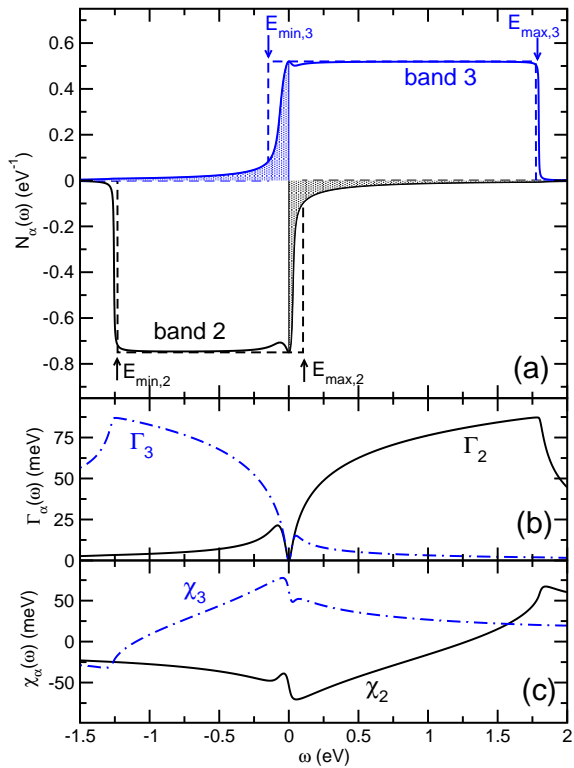}
\caption{(color online). Panel (a): DOS of the hole band 2 and of the
  electron band 3 (solid lines), displayed along with the bare DOS (dashed
  lines). To better resolve the two bands we used a negative $y$ axis for the
  DOS of the band 2. The shaded areas give the number of carriers $n_\a$ in each
  band at $T=0$, according to  Eqs.\ \pref{defdensh}-\pref{defdense}. 
Panels (b) and (c): energy dependence of the imaginary and of the real
part of the self-energy for the same bands.
Note that $\G_2(\o)$ is different from zero in an energy
range corresponding to the DOS of the band 3, and viceversa. }
\label{f-selfenergy}
\end{figure}

To better understand this issue,
we plot in Fig.\ \ref{f-selfenergy} both the DOS
and the self-energy for the band 2 and 3 over a larger energy scale,
which allows us to show also that similar effects occur
for both electron- and hole-like bands.
In panel Fig.\ \ref{f-selfenergy}b we compare the corresponding
imaginary parts of the self-energy $\Gamma_\a(\o)$.  It is important
to notice here that in the hole-like band $\Gamma_2(\o)$ has a finite
support that coincides approximately with the DOS of the electronic
band $N_3(\o)$.  The same occurs for $\Gamma_3(\o)$, which is
controlled by the support of $N_2(\o)$.  This is a direct consequence
of the interband character of the interaction, where the energy
support of $\Gamma_\a(\o)$ is determined by the DOS of the band $\b$
to which the band $\a$ is coupled.  This situation has drastic effects
on the resulting DOS. Indeed, as shown in Fig.\ \ref{f-selfenergy}a,
the incoherent states induced by $\Gamma_\a(\o)$ are thus created
mainly {\em outside} the energy range of the band $\a$, namely {\em
  below} the band bottom for the electron dispersions and {\em below}
the band top for the hole ones, and they are responsible for the long
tails in the corresponding DOS.  It is worth stressing here that such
tails are {\em not} due to a strong-coupling effect, but simply to a
predominant interband nature of the interaction.

\section{Optical conductivity and spectral weight}
\label{s-opt}

In the previous Section we have seen how the Fermi-surface shrinking
generated by a retarded interaction is not directly reflected in a
substantial reduction of the charge carrier density $n_\a$ for the
corresponding band, although the total number of coherent states
$\tilde{n}_\a$ is strongly affected by that shrinking.  As
we shall see in the present Section, such different behavior
can be properly detected in the analysis of the optical properties.
In particular we show that the optical sum rule, which is directly
related to total number of carrier $n_\a$ in each band, is almost
unaffected by the Fermi-surface shrinking, whereas the DC conductivity
and the spectral weight in the coherent Drude peak scale with
$\tilde{n}_\a$, so that  they are strongly reduced by the Fermi-surface
shrinking, with a transfer of spectral weight to the intraband
incoherent processes.

To address this issue we compute explicitly the optical conductivity
in the presence of the retarded interband interaction.  Since for a
$\bk$-independent self-energy vertex corrections vanish,
$\sigma(\omega)$ can be computed in the simple-bubble approximation.
We can write thus $\sigma(\omega)=\sum_\a \sigma_\a(\omega)$ , where
\bea
\lb{sigma}
\sigma_\alpha(\omega)&=&-\frac{2\pi e^2}{\hbar } 
\int_{-\infty}^\infty dz \frac{f(z-\mu+\o)-f(z-\mu)}{\omega}\nn\\
&\times &\int \frac{d^2\bk}{(2\pi)^2} v^2_{{\bf k},\a}
A_\a(\e_{{\bf k},\a},z+\o)A_\a(\e_{{\bf k},\a},z),
\eea 
where $v_{{\bf k},\a}=(1/\hbar)\pd \e_{{\bf k},\a}/\pd k_x$ is the
quasiparticle velocity and the spectral function can be written
explicitly as:
\be
\lb{spectral}
A_\a(\e,\o)=\frac{1}{\pi}\frac{\Gamma^{\rm qp}_\a(\o)}{[\o-\e-\chi_\a(\o)]^2+[\G_\a^{\rm qp}(\o)]^2}.
\ee
Here we considered also a finite contribution of disorder to the total quasi-particle
scattering rate, $\G_\a^{\rm qp}(\o)\equiv \G_\a(\o)+\G_0$. Since as
$T\ra 0$ $\G_\a(0)\ra 0$ we choose $\G_0=10$ meV by estimating the
approximate width of the low-energy optical spectra 
from Ref.\ \cite{basov} (see also Eq.\ \pref{deftau}-\pref{drude} below). 
By converting the $\bk$ integration in an energy integration and
considering the parabolic hole-like and electronic-like
dispersions described in Eqs. (\ref{disph})-(\ref{dispe}),
we obtain for the hole bands ($\a=1,2$)
\bea
\sigma_\alpha(\omega)&=&- \frac{e^2}{\hbar} 
\int_{-\infty}^\infty dz \frac{f(z-\mu+\o)-f(z-\mu)}{\omega}\nn\\
&\times&\int_{E_{\rm min}}^{E_{\rm max}}d \epsilon (E_{{\rm max},\a}-\e)
A_\a(\e,z+\omega)A_\a(\e,z),\nn\\
\lb{sigma_h}
\eea 
while for the electron bands ($\a=3,4$)
\bea
\sigma_\alpha(\omega)&=&- \frac{e^2}{\hbar} 
\int_{-\infty}^\infty dz \frac{f(z-\mu+\o)-f(z-\mu)}{\omega}\nn\\
&\times&\int_{E_{\rm min}}^{E_{\rm max}}d \epsilon (\e-E_{{\rm min},\a})
A_\a(\e,z+\omega)A_\a(\e,z).\nn\\
\lb{sigma_e}
\eea 
\begin{figure}[t]
\includegraphics[scale=0.3,angle=0,clip=]{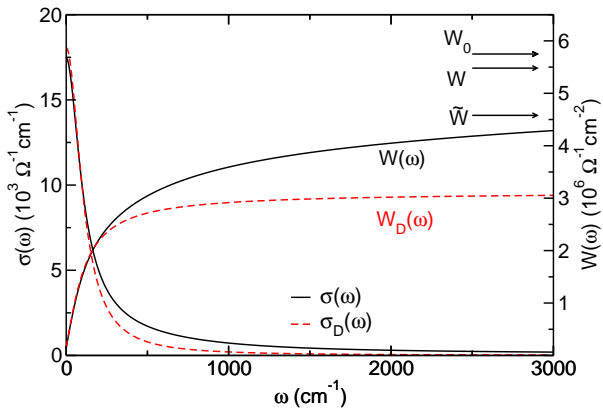}
\caption{(color online)
Left axis: Frequency dependence of the optical conductivity at
$T=5$ K  as obtained from \pref{sigma_h}-\pref{sigma_e} (solid line),
and  low-energy  Drude-like peak as obtained
from Eq. (\ref{drude}) (dashed line).
The small discrepancy for the dc values stems from
the use of the approximate formulas \pref{shappe}-\pref{shapph} in the Drude term. To
permit a simpler comparison with the experiments we divided the two-dimensional results by  
the interlayer distance $d=8.5$ \AA. 
Right axis: integrated spectral weight $W(\omega)$, compared to the
integral of the Drude part only, $W_{\rm D}(\o)$. 
The arrows mark 
the bare value $W_0$, the asymptotic value $W$ \pref{common.rule}
in the interacting case, and the value $\tilde W$ corresponding to the
reduced Fermi
surfaces.} 
\label{f-opt}
\end{figure}
The resulting optical conductivity $\sigma(\o)$ at low temperature is
shown in Fig.\ \ref{f-opt} for the same coupling value $V=0.46$ eV
used in Ref.\ \onlinecite{cappelluti_dhva} to reproduce the Fermi-surface
shrinking measured by de Haas-van Alphen experiments in LaFePO.  Also
shown in the same plot is the integrated optical spectral weight
$W(\omega)=\int_0^{\omega} d\o'\sigma(\omega')$ as a function of the
cut-off frequency. When the integration frequency goes to infinity the
total spectral weight is given by the optical sum
rule, that for a parabolic-band approximation as the one we are using
here simply reduces to\cite{review_benfatto,review_basov}
\be 
\lb{common.rule} 
W
= \int_0^\infty d\o \sigma(\o)=
\frac{\pi e^2}{2}\sum_\a \frac{n_\a}{m_\a}.
\ee
This asymptotic value is marked in Fig.\ \ref{f-opt} by a solid arrow,
along with the sum-rule value $W_0$ in the absence of the coupling to
spin fluctuations. As we can see, the discrepancy between $W$ and
$W_0$ results to be of order of 5 \%, pointing out that the effect of
the Fermi-surface shrinking induced by the many-body interaction is
quite small in the total sum rule. However, by direct inspection of
Fig.\ \ref{f-opt} one can also see that even at frequencies $\o\sim
3000$ cm$^{-1}$, corresponding to about $20\o_0$, the collected spectral
weight $W(\o)$ still differs by about $20\%$ from its asymptotic limit
\pref{common.rule}, and it is near instead to the spectral weight
corresponding to a rigid-band shift, i.e.
$\tilde{W}=\pi e^2 \sum_\a \tilde n_\a/2m_\a$.
Such a lack of saturation of $W(\o)$ over
frequency scales much larger than the typical boson energy $\o_0$ is
unusual for single-band systems,\cite{review_carbotte,karakozov} and
indicates that for the multiband case discussed here a large spectral
contribution is allocated at relatively high energies.

To enlighten this issue and its relation to the Fermi-surface
shrinking let us first evaluate the dc conductivity in the zero
temperature limit, where we can write in Eqs.\
\pref{sigma_h}-\pref{sigma_e} $ \lim_{\o\ra
  0}[{f(z-\mu+\o)-f(z-\mu)}]/{\omega}\approx -\d(z-\mu)$. By performing the
remaining integration over $\e$ analytically we get thus:
\bea
\sigma_\a^{\rm dc}&=&\frac{e^2}{2\pi^2\hbar}
\left\{ 1+\frac{|E_{\rm min,\a}|-\chi_\a(0)}{\G_\a^{\rm qp}(0)}
\left[ \arctan\frac{E_{\rm max,\a}+\chi_\a(0)}{\G_\a^{\rm qp}(0)}\right.\right.\nn\\
&+&\left.\left. \arctan\frac{|E_{\rm min,\a}|-\chi_\a(0)}{\G_\a^{\rm qp}(0)}\right]\right\}
\eea
for the electrons and
\bea
\sigma_\a^{\rm dc}&=&\frac{e^2}{2\pi^2\hbar}
\left\{ 1+\frac{E_{\rm max,\a}+\chi_\a(0)}{\G_\a^{\rm qp}(0)}
\left[ \arctan\frac{|E_{\rm min,\a}|-\chi_\a(0)}{\G_\a^{\rm qp}(0)}\right.\right.\nn\\
&+&\left.\left. \arctan\frac{E_{\rm max,\a}+\chi_\a(0)}{\G_\a^{\rm qp}(0)}\right]\right\}
\eea
for the holes. For $T\ra 0$ the only source of damping is due to
disorder, so that $\G^{\rm qp}(0)$ reduces to
impurity scattering $\G_0$. Since $\G_0$ is usually much smaller than the
distance of the both/top of the band from the Fermi level [including also
the shift $\chi(0)$], the above expressions simplify considerably. Indeed,
using the expressions
\pref{kfh},\pref{kfe}, \pref{nshift},
we obtain for the hole bands ($\a=1,2$)
\be
\lb{shapph}
\sigma^\a_{\rm dc}
=\frac{e^2}{2\pi \hbar \G^{\rm qp}(0)}[E_{{\rm max},\a}+\chi_\a(0)]=\frac{
  \tilde n_\a e^2 \tau_{{\rm tr},\a}}{m^*_\a},
\ee
and for the  electron bands ($\a=3,4$)
\be
\lb{shappe}
\sigma^\a_{\rm dc}
=\frac{e^2}{2\pi \hbar \G^{\rm qp}(0)}[|E_{{\rm min},\a}|-\chi_\a(0)]=\frac{
  \tilde n_\a e^2 \tau_{{\rm tr},\a}}{m^*_\a},
\ee
where we introduced the effective mass $m^*_\a=(1+\l_\a)m_\a$
and the transport scattering time 
\be
\lb{deftau}
\tau_{{\rm tr},\a}^{-1}=\frac{2\G^{\rm qp}_\a(0)}{\hbar (1+\lambda_\a)}.
\ee
As it is evident from
Eqs. (\ref{shapph})-(\ref{shappe}), even though the total carrier numbers $n_\a$ are
almost unaffected by the shrinking of the Fermi areas (see Fig.\
\ref{f-dens-vs-chi}), the dc conductivity is instead controlled by the
coherent carrier density $\tilde{n}_\a$.  More generally, one can
generalize this result to finite frequency to get an approximate
expression of the coherent Drude peak $\sigma^{\rm D}(\omega)$ in the
presence of the many-body interaction as
\be
\lb{drude}
\sigma^{\rm D}_\a(\o)=\frac{\tilde{n}_\a e^2\tau_{{\rm tr},\a}}{m^*_\a}
\frac{1}{1+\left(\o \tau_{{\rm tr},\a}\right)^2}.
\ee
The coherent Drude peak described in Eq. (\ref{drude}) is also shown
in Fig. \ref{f-opt}, along with its integrated spectral weight $W_{\rm
  D}$ on the right-side scale. The comparison with $W(\o)$ shows that
$W_{\rm D}(\o)$ saturates quite rapidly to its asymptotic value $\pi e^2\tilde
n/2m^*$, while $W(\o)$ steadily increases in this range of energies,
showing that all the remaining
spectral weight needed to recover the value \pref{common.rule} must
originate from incoherent processes at higher frequencies. Note that
this result has to be contrasted with the standard infinite-band case,
where the real part of the self-energy vanishes and the
spectral-weight distribution is only due to the mass renormalization,
so that the coherent Drude part $W_{\rm D} \propto n_\a/m^*_\a$ (see
Fig.\ \ref{f-skt}a).\cite{review_basov} On the contrary, here the many-body interaction
gives rise to an additional Fermi-surface shrinking which is reflected
in a further reduction of the coherent spectral weight $\tilde{W} \propto
\tilde{n}_\a/m^*_\a$ (see Fig.\ \ref{f-skt}c).

\section{Temperature dependence}

In the previous section we have seen that the many-body interband interaction
plays a  non trivial role on the spectral properties of
systems with strongly particle-hole asymmetric bands, and
in particular in pnictides.
This is pointed out for instance in the coherent contribution
to the optical conductivity described in Eq. (\ref{drude}).
In infinite bandwidth systems,
assuming the bosonic spectrum $B(\Omega)$ to be weakly temperature
dependent, the main temperature dependence
of the optical conductivity
in Eq. (\ref{drude}) is through the parameter $\tau_{{\rm tr},\a}(T)$
[or equivalently $\Gamma^{\rm qp}_\a(T)$]
whereas $\tilde{n}=n$ is temperature independent.
This scenario has to be contrasted with the case of a multiband system
with strong
particle-hole asymmetry, where the real part of the
self-energy $\chi_\a(0)$, which gives rise to the
Fermi surface shrinking, is itself temperature dependent
and it is reflected in a coherent charge concentration $\tilde{n}_a$
which acquires a finite temperature dependence.
Such a case will be analyzed in the Section \ref{ssI}, whereas
in Section \ref{ssII} we consider the additional effects due to a temperature-dependent
bosonic spectrum, as suggested by a number of experiments in pnictides.
In this case the dimensionless electron-boson coupling $\lambda_\a$
and the low-energy mass renormalization $m_\a^*/m_\a=1+\lambda_\a$
acquire their own temperature dependence, which can compete with
the previously discussed effects.
As we shall see,
transport and optical properties can present quite different
temperature behaviors in the two cases.

\subsection{Temperature-independent bosonic spectrum}
\label{ssI}

\begin{figure}[t]
\includegraphics[scale=0.3,angle=0,clip=]{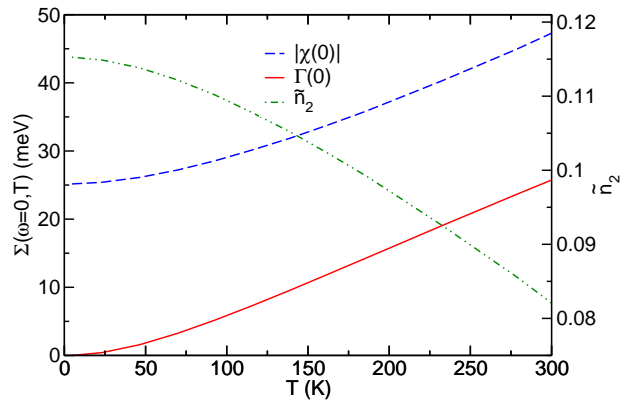}
\caption{(color online) Left axis: temperature dependence
of the real and imaginary part of the self-energy at zero frequency
for band 2. Right axis: temperature
dependence of the coherent carrier density $\tilde{n}_2$
obtained from the Fermi-surface area according Eq. \pref{nshift}.} 
\label{f-sigmaT}
\end{figure}

We consider first the case where
the bosonic spectrum is not itself temperature dependent.
In this case the formalism is the same as we discussed so far, and we
should only account for the temperature evolution of the
self-energy. In Fig. \ref{f-sigmaT} we show
as an example the temperature dependence
of the real and imaginary part of the self-energy at zero frequency
for band 2. As we can see,
the absolute value of both quantities increases as the temperature
increases, due to the thermal excitations of the bosonic mode. 
To get some analytical insight into these results one can employ the
first-order expansion for the self-energy, computed using the non-interacting
Green's function in Eq.\ \pref{self}. The corresponding expression for 
$\chi_\a(0,T=0)$ has been derived in Ref.\
\onlinecite{cappelluti_dhva} in the case of an Einstein bosonic spectrum, 
$B(\O)=(\o_{\rm E}/2) \d(\O-\o_{\rm E})$. By performing the same calculation
at finite $T$ one can see that the most relevant temperature
corrections are related to the thermal excitation of the boson:
\bea
\lb{chit}
\chi^E_\a(0,T)
&\approx&
 -\frac{\o_{\rm E}}{2}\sum_\b \l_{\a\b} \ln
\left|\frac{E_{{\rm max},\b}-\mu} {E_{{\rm min},\b}-\mu} \right|
\nonumber\\
&&\times
\left[ 1+\frac{2\o_{\rm E}}{ E_{c,\b}}b(\o_{\rm E}/T)\right],
\eea
where $E_{c,\b}={\rm min} (E_{{\rm max},\b}, E_{{\rm min},\b})$
is the nearest band edge, and $b(x)=1/[\exp(x/T)-1]$ is the Bose
function. In the case of a spin-fluctuation spectrum the direct comparison
with the results in Fig.\ \ref{f-sigmaT} shows that 
a similar expression holds, provided the identification
of the characteristic energy scale
$\o_{\rm E}\approx 2\o_0$.
The imaginary part of the self-energy at zero frequency can be
also computed exactly in the same approximation
for temperatures $T\ll E_c$:
\be
\lb{g0t}
\G^E_\a(0,T)\approx -\pi T \sum_\b \l_{\a\b}.
\ee
An analytical results can be derived for $\G_\a(0,T)$
also for the spin-fluctuation spectrum, and we obtain
\be
\lb{g0tsf}
\G_\a(0,T)\approx -2T\arctan \frac{T}{\o_0} \sum_\b \l_{\a\b},
\ee
which is in good agreement with the results of Fig.\ \ref{f-sigmaT}.

For what concerns the optical spectrum, the thermal excitation of the
bosonic mode has then two consequences: ($i$) a further reduction of
the Fermi-surface areas, encoded in the decrease of $\tilde{n}_\a(T)$ with
temperature, as shown in Fig.\ \ref{f-sigmaT}; ($ii$) a progressive
broadening of the optical conductivity features, due to the increase
of $\G_\a(T)$. While the latter effect is customary also in the
single-band infinite-bandwidth case, the former one is peculiar of
multiband systems with interband-dominating interactions, and leads to
an unusual temperature dependence of the partial optical sum rule
$W(\omega_c)$.  Indeed, as we have seen in Section \ref{s-opt}, since
the Fermi-surface shrinking gives rise to a transfer of spectral
weight from the low-energy Drude peak $\propto \tilde{n}$ to the
incoherent structure extending up to high energies, a finite
temperature dependence of $\tilde{n}$ reflects directly in a sizable
temperature-dependent depletion of $W(\omega_c)$ when a finite cut-off
$\omega_c$ is employed.

\begin{figure}[t]
\includegraphics[scale=0.3, angle=0,clip=]{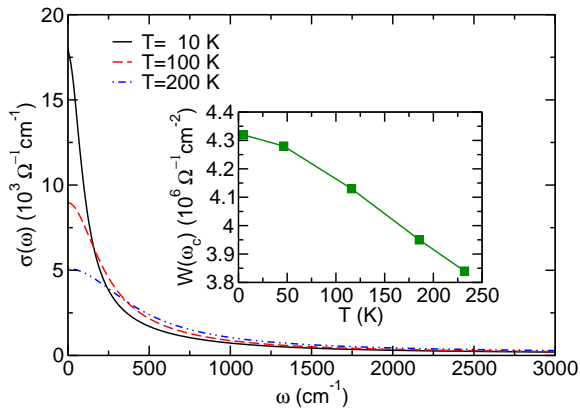}
\caption{(color online) Optical conductivity
for three representative temperatures $T=10,100,200$ K
using a temperature-independent bosonic spectrum $B(\Omega)$
and an impurity scattering rate $\G_0=10$ meV.
Inset: temperature dependence of the
integrated spectral weight $W(\omega_c)$ with $\omega_c=3000$ cm$^{-1}$.
}
\label{f-opt-T}
\end{figure}

In Fig.\ \ref{f-opt-T} we show the evolution of the optical
conductivity for three different temperatures, using again 
the multiband scattering matrix
$\lambda_{\a\b}$ and the bosonic spectrum estimated in
Ref. \onlinecite{cappelluti_dhva} to reproduce the Fermi-surface shrinking 
of LaFePO at $T\approx 10$ K. Here we assume the all these parameters
are temperature independent and we also use a constant impurity
scattering rate $\Gamma_0=10$ meV.  The corresponding variation of the
spectral weight $W(\omega_c)$ as a function of the temperature is
reported in the inset, showing an overall reduction of about 10 \%.
Note that such a variation is an order of magnitude larger than the
one expected in a single-band system for the same coupling value, even when
the presence of a finite bandwidth\cite{karakozov,benfatto_prb06} and
a finite cut-off\cite{norman_prb07,chubukov_prb10} are taken into
account.  Quite remarkably this temperature variation is as large as
the one observed in cuprate superconductors, where it has been
interpreted as a consequence of strong
correlations.\cite{review_benfatto,review_carbotte,review_basov} In
our case however this effect comes primarily from the temperature
dependence of the coherent carriers number, that is relatively large
even when the coupling to the bosonic mode is not particularly strong.

\subsection{Temperature-dependent bosonic spectrum}
\label{ssII}

\begin{figure}[t]
\includegraphics[scale=0.3,angle=0,clip=]{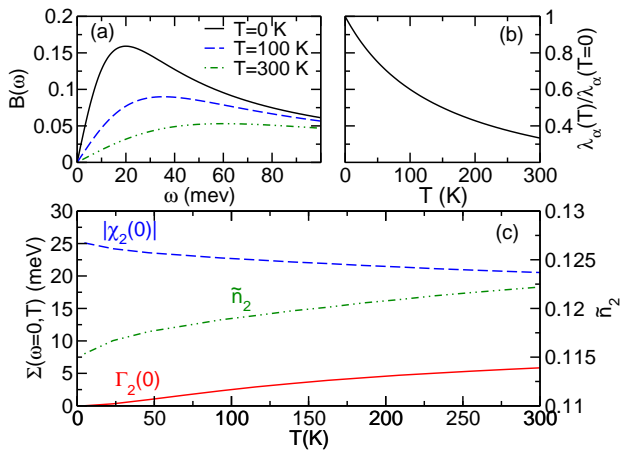}
\caption{(color online) (a) Temperature evolution of the bosonic
  spectrum according to Eqs.\ \pref{spectrum}-\pref{omt}. (b)
  Temperature evolution of the dimensionless coupling constant in each
  band. (c) Corresponding temperature dependence of $\chi_2(0)$ and
  $\Gamma_2(0)$ (left axis) and of $\tilde{n}_\a$ (right axis) in band
  2 using the temperature-dependent bosonic spectrum $B(\Omega)$ of
  panel (a).}
\label{f-sigmaT-exp}
\end{figure}

\begin{figure}[t]
\includegraphics[scale=0.3, angle=0,clip=]{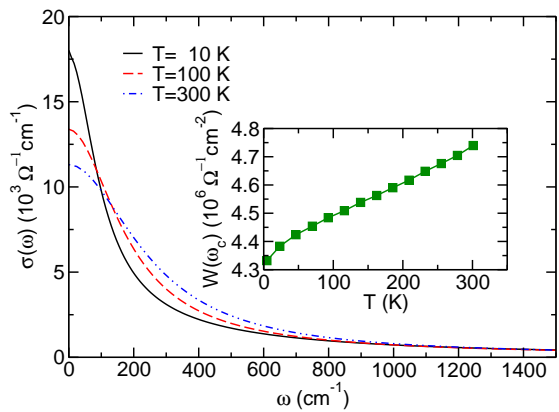}
\caption{(color online).  Optical conductivity for three
  representative temperatures using the temperature-dependent bosonic
  spectrum $B(\Omega)$ shown in Fig.\ \ref{f-sigmaT-exp}a. Here we
  used a constant impurity scattering rate $\G_0=10$ meV.  Inset:
  temperature dependence of the integrated spectral weight
  $W(\omega_c)$ with $\omega_c=3000$ cm$^{-1}$.}
\label{f-opt-T-exp}
\end{figure}

As we have seen in Sec. \ref{s-opt}, the low-energy part of the
optical spectrum resembles a renormalized Drude model, with an inverse
scattering time $\tau_{\rm tr}^{-1}$ proportional to the imaginary
part of the self-energy. As a consequence, if the boson spectrum is
constant in temperature the broadening of the optical conductivity as
the temperature increases is directly proportional to the increase of
$\G_\a(0)$, as shown for instance in the case of Fig.\ \ref{f-opt-T}.
With this choice, however, we get an optical width of the Drude-like
low-frequency part at room temperature of about 600 cm$^{-1}$, much
larger than then scattering rate $\tau_{\rm tr}^{-1}\simeq 320$
cm$^{-1}$ that can be fitted from the available experimental data at
$T=300$ K.\cite{basov} This observation suggests thus that the
increase of $\G_\a(0)$ with temperature is smaller than predicted by
Eq.\ \pref{g0t}. To account for this discrepancy we investigate here
the possibility that the bosonic spectrum is itself temperature
dependent, as it has been suggested also by a recent analysis of
optical spectra in 122 compounds.\cite{carbotte_prb10} In particular,
in Ref.\ \onlinecite{carbotte_prb10} it has been shown that the bosonic
spectrum extracted from optics hardens as $T$ increases, while its
total spectral weight, encoded in the dimensionless couplings
$\l_{\a\b}$, decreases. This trend is also consistent with direct
measurements of the spin-fluctuations spectrum by means of neutron
scattering in 122 compounds.\cite{inosov_natphys10} Following
Ref. \onlinecite{inosov_natphys10}, we model thus such behavior by
assuming a temperature-dependent bosonic spectrum of the form:
\be
\lb{spectrum}
B(\O)= \frac{h(T)}{\pi}  \frac{\bar\o_0\o}{\o^2+\bar\o_0^2 }
\ee
where
\be
h(T)=\frac{T_\theta}{T+T_\theta},
\ee
and
\be
\lb{omt}
\bar\o_0(T)= \frac{\o_0}{h(T)}.
\ee
We chose the value of $T_{\theta}=150$ K in order to reproduce
the experimental value of the Drude peak in the optical data of LaFePO at
$T=300$ K.\cite{basov}
The evolution of the spectrum $B(\Omega)$ for three representative
temperatures $T=0,100,300$ K and the corresponding
temperature dependence of the coupling constant are shown in 
Fig. \ref{f-sigmaT-exp}a and \ref{f-sigmaT-exp}b, respectively. 
Focusing for simplicity once again on band 2, we plot in panel (c) the
resulting real and imaginary parts of the self-energy $\chi_2(0)$,
$\Gamma_2(0)$.  As one can see, for this choice of parameters the
reduction of $\lambda_2(T)$ is strong enough to result in a {\em
  decrease} of |$\chi_2(0)$| with the temperature and in a smaller
increase of $\Gamma_2(0)$.  As a consequence, also the temperature
evolution of the optical conductivity, shown in Fig.\
\ref{f-opt-T-exp},
is markedly different than before.
In particular, the reduction of |$\chi_\a(0)$| with the temperature
is reflected now in an {\em increase} of $\tilde n_\a(T)$ (see Fig.\
\ref{f-sigmaT-exp}c).
The spectral weight moves thus
progressively toward the low-energy coherent part of the spectrum, leading
also to an effective increase of the spectral weight $W(\o_c)$ integrated in a fixed
frequency windows, in sharp contrast to what expected in ordinary
single-band (interacting and non-interacting)
systems.\cite{review_benfatto,review_basov,review_carbotte}

\section{Comparison with the experiments}

Even though a detailed comparison with the experimental optical spectra is
beyond the scope of the present paper, we would like to comment on the
possible consequences of our findings for the physics of pnictides.  A
first issue concerns the role of the cut-off in the estimate of both the
absolute value of the spectral weight and its temperature dependence.  From
the experimental point of view, a finite cut-off $\o_c$ is always employed
in the estimate of the sum rule to avoid inclusion of interband
transitions,
so that $W_{\rm exp}\sim
W(\o_c)$.  In pnictides the choice of $\omega_c$ is dictated by the
observed threshold to interband transitions, that occurs usually around
$\omega_c=2000-3000$ cm$^{-1}$.  This is indeed the case in Ref.\
[\onlinecite{basov}], where the experimental estimate $W_{\rm exp}$ for
LaFePO includes only processes up to a cut-off frequency of $\o_c=3000$
cm$^{-1}$. As we have shown in Fig.\ \ref{f-opt}, the calculated spectral
weight collected up to this same cut-off is about 20$\%$ of its asymptotic
value $W$, which is instead almost unchanged (within $\sim 5\%$) from the
bare value $W_0$ (we remind here that in our model $W_0$.
While comparing with the experiments we must recall that
our starting model has renormalized DFT bands, so that $W_0=0.5 W_{\rm
  DFT}$. As a consequence, after inclusion of the low-energy spin
fluctuations we estimate overall $W(\o_c)\simeq 0.4 W_{\rm DFT}$, i.e. the
main source of spectral-weight reduction still comes from correlation
effects. We observe however that for the range of parameters used here
$W(\o_c)$ is very near to the spectral weight $\tilde{W}$ that one would
obtain from a rigid-band shift.  The reason is that, as we have seen in
Sec. II, the incoherent part of each spectral function extends up to energy
scales of the order of the bandwidth of the coupled band, i.e. over energy
scales much larger than the typical boson energy $\o_0$. As a consequence,
even though there is no theoretical sum rule relating $W(\o_c)$ to
$\tilde{W}$, on the energy scales considered in the experiments this
could be a good approximation for it. Note however that the situation could
be slightly different in the 122 compounds, where the most populated band
is also the less coupled to spin fluctuations,\cite{benfatto_prb09} making
it difficult to predict a-priori in a precise quantitative way the 
relevance of the shrinking on the absolute value
of optical sum rule with respect to DFT. 

A more direct effect of the Fermi-surface shrinking on the optical data
is instead the temperature dependence of sum rule. Indeed, as we
discussed in the previous Section, the temperature dependence of
$W(\o_c)$ is strongly affected by the temperature evolution of the
coherent charge density $\tilde n_\a$, which in turn is an indirect
probe of the temperature evolution of the spin-fluctuation
spectrum. Quite interestingly, recent optical measurements in both
carrier-doped\cite{degiorgi,dressel_prb10} and
isovalent-substituted\cite{dressel_cm10} 122 compounds show that
$W_{\rm exp}$ increases with increasing temperature, in contrast to
any ordinary behavior in single-band
system.\cite{review_benfatto,review_basov,review_carbotte} Within our
scenario this suggests that the coupling to the spin fluctuations is
actually temperature dependent, leading to an increase of $\tilde
n_\a$ as the temperature increases, see Fig.\ \ref{f-sigmaT-exp}c, and
to a progressive population of the low-energy optical spectrum, see
Fig.\ \ref{f-opt-T-exp}. It is worth noting that our
approach completely neglects the role of optical ({\bf q}=0) interband
transitions, that are presumably less relevant than the single-band
transitions in the range of frequencies considered here.
However, a detailed analysis of the role of these interband transitions and of
their possible relevance in the experiments is at the moment an
open question that deserves further theoretical
investigation.

A third outcome of our results is related to the
possible observation of these anomalous temperature behaviors in the
dc transport properties.  Indeed, from the approximate expressions
\pref{shapph}-\pref{shappe} for the dc conductivity it is clear that as the
temperature increases both $\tilde n_\a(T)$ and $\G_a(T)$ contribute to
the temperature dependence of the dc conductivity.  In other words for
a multiband system with dominant interband interactions one cannot
simply attribute the $T$ dependence of the resistivity to
scattering processes, since also the density of coherent carriers can
have a non trivial temperature dependence. This observation could shed
new light also on the comparative analysis of dc and Hall
conductivities carried out in Refs.\
[\onlinecite{rullier_prl09,rullier_prb10}], where part of the anomalous
behavior reported there could be accounted for by relaxing the
assumption of a constant density of carriers with the temperature.

\section{Conclusions}

In this paper we analyzed systematically the behavior of the
optical-conductivity sum rule in a multiband system with dominant
interband interactions, as it is the case in pnictide
superconductors. Due to the strong particle-hole asymmetry of these
systems, the coupling to a collective mode leads to a many-body
induced band shift and to a corresponding shrinking of the Fermi-surface
areas. In this context, we have shown however that the number of
particles in each band does not scale with the Fermi areas, but it is
almost unchanged.  Such behavior has been explained in terms of a
transfer of the spectral weight from the coherent states related to
the reduced Fermi areas to the unoccupied part of the non-interacting
bands.  As we have discussed in the present paper, the effects of such
redistribution of spectral weight on the optical spectrum are quite
different from what expected in single-band systems.  In particular, we
have shown that only a fraction $\sim \tilde n/m^*$ of the total
weight $\sim n/m$ is collected in the coherent low-energy part of the
spectrum, while the rest is recovered as higher-energy incoherent
processes. As a consequence, the unavoidable use of a finite cut-off
$\o_c$ in the experimental determination of the optical sum rule is
expected to have a much stronger effect than in the single-band
case. There are two main consequences: (i) $W_{\rm exp}$ usually
underestimates the theoretical sum-rule value $W$ in a percentage that
cannot be established a-priori, since it depends on the parameter
values. In the case of LaFePO discussed here, $W_{\rm exp}$ turns out
to be about $20\%$ smaller than $W$. As a consequence, if one assumes
that $W$ is already strongly reduced ($\sim 50\%$)  by
correlations,\cite{georges,vollhardt}  the
many-body effect due to spin fluctuations is a small fraction ($\sim
10\%$) of the DFT value $W_{DFT}$; (ii) the temperature evolution of
the optical sum rule turns out to be an indirect measurement of the
temperature evolution of the Fermi-surface areas.  For instance, a temperature
dependence of the spin-fluctuation spectrum could suggest that the
$\omega$-integrated spectral weight $W(\o_c)$ could {\em increase}
with temperature, in sharp contrast to the case of interacting
single-band systems, as for example cuprate superconductors. All these
findings pose strong constraints on the analysis of optical data in
Fe-based materials, showing once more that the peculiar semimetal
and multiband character of these systems forces us to revise 
standard paradigms appropriate for single-band correlated materials.

\section{Acknowledgements}
We thank C. Castellani, L. Boeri and L. Ortenzi for useful discussions. 
This work has been supported in part by the Italian
MIUR under the project PRIN 2007FW3MJX and PRIN 2008XWLWF9.


\begin{thebibliography}{99}
\bibitem{note-edge} 
It is clear that this value does not represent the true band edge as given
in DFT, but an effective one needed to guarantee the 
total number of states.
\end{thebibliography}
\end{document}